\documentclass[aps,prl,twocolumn,superscriptaddress,showpacs]{revtex4}

\usepackage{amsmath}
\usepackage{graphicx}
\usepackage{calc}
\usepackage{bm}

\begin{document}
\title{Classical effects in the weak-field magnetoresistance of InGaAs/InAlAs quantum wells}
\author{M.~Yu. Melnikov}
\affiliation{Institute of Solid State Physics, Chernogolovka, Moscow District 142432, Russia}
\author{A.~A. Shashkin}
\affiliation{Institute of Solid State Physics, Chernogolovka, Moscow District 142432, Russia}
\author{V.~T. Dolgopolov}
\affiliation{Institute of Solid State Physics, Chernogolovka, Moscow District 142432, Russia}
\author{G. Biasiol}
\affiliation{IOM CNR, Laboratorio TASC, 34149 Trieste, Italy}
\author{S. Roddaro}
\affiliation{NEST, Istituto Nanoscienze-CNR and Scuola Normale Superiore, 56127 Pisa, Italy}
\author{L. Sorba}
\affiliation{NEST, Istituto Nanoscienze-CNR and Scuola Normale Superiore, 56127 Pisa, Italy}

\begin{abstract}
We observe an unusual behavior of the low-temperature magnetoresistance of the high-mobility two-dimensional electron gas in InGaAs/InAlAs quantum wells in weak perpendicular magnetic fields. The observed magnetoresistance is qualitatively similar to that expected for the weak localization and anti-localization but its quantity exceeds significantly the scale of the quantum corrections. The calculations show that the obtained data can be explained by the classical effects in electron motion along the open orbits in a quasiperiodic potential relief manifested by the presence of ridges on the quantum well surface.
\end{abstract}
\maketitle

Much interest has been attracted recently by the properties of the two-dimensional (2D) electron gas in InGaAs/InAlAs quantum wells, like the strong spin-orbit Rashba interaction, large Lande $g$-factor, and small effective mass. The strong anisotropy of the mobility in (100) quantum wells with In content $\geq 0.75$ was observed in a number of publications \cite{kav,gold,rich,gozu,lohr,rosini}. In the work of Ref.~\cite{biasol}, it was shown that the anisotropy originates largely from the modulation of In content that leads to the modulation of conduction band bottom. A relief on the sample surface was observed, reflecting the modulation of In content. The authors of Ref.~\cite{biasol} also suggested a way to simultaneously increase the electron mobility and reduce the mobility anisotropy by introducing an InAs layer into the center of the quantum well. As long as the fluctuations of In content are impossible in the binary composition, the degree of disorder in the quantum well should decrease. Indeed, the mobility anisotropy was found to decrease as a result of improving the growth structure. However, it remained unclear whether the residual anisotropy in the transport properties of InGaAs/InAlAs quantum wells is the only manifestation of the built-in growth anisotropy.

In this paper, we study the low-temperature magnetotransport properties of the 2D electron gas in InGaAs/InAlAs quantum wells with an InAs layer in weak perpendicular magnetic fields. The observed magnetoresistance is unusual in that it is qualitatively similar to that expected for the weak localization and anti-localization but its quantity exceeds significantly the scale of the quantum corrections. The calculations show that the obtained data can be explained by the classical effects in electron motion along the open orbits in a quasiperiodic potential relief manifested by the presence of ridges on the quantum well surface. The results indicate that the likely origin for the potential relief is the residual modulation of In content in the quantum well.

\begin{figure}[b]
\scalebox{0.6}{\includegraphics[width=\columnwidth]{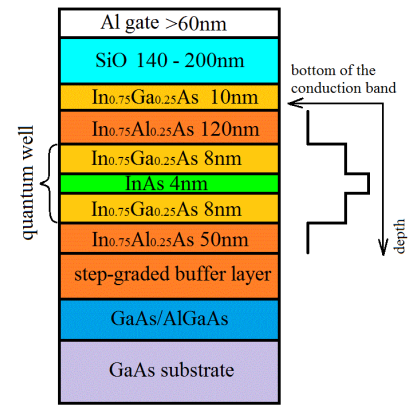}}
\caption{Growth structure of the wafer. Also shown is the conduction band diagram.}
\label{fig1}
\end{figure}

Measurements were performed in a dilution refrigerator in a range of temperatures between 0.03 and 1.2~K on Hall bar samples with [011] and [01-1] orientations fabricated based on structures shown schematically in Fig.~\ref{fig1}. After etching a mesa and evaporating Ni/Au contacts to the 2D electron gas, a SiO layer was evaporated on the wafer surface and then an Al gate was evaporated on the top of SiO. At zero gate voltage, the low-temperature mobility was about 48~m$^2$/Vs at electron density $n_s\simeq 5\times 10^{11}$~cm$^{-2}$ for currents (and Hall bars) in both [011] and [01-1] directions. The resistance was studied using four-terminal measurements on four samples (two samples for each orientation).

\begin{figure}
\scalebox{0.7}{\includegraphics[width=\columnwidth]{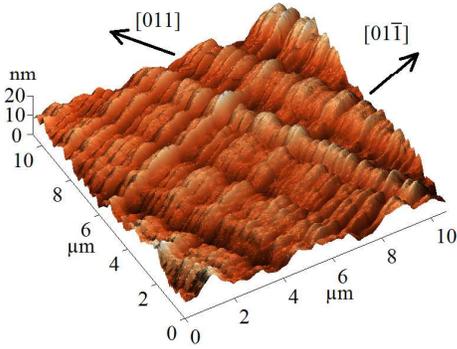}}
\caption{Atomic-force microscope image of the wafer surface.}
\label{fig2}
\end{figure}

The surface of the wafer was studied using an atomic-force microscope. A typical scan of an area $10\times 10$~$\mu$m$^2$ is displayed in Fig.~\ref{fig2}. Ridges are discernible along [011] and [01-1] directions with characteristic height 10~nm. The ridges along the [011] direction are quasiperiodic in the [01-1] direction with characteristic period 1.5~$\mu$m. The other set of ridges along the [01-1] direction is distinguished by a smaller characteristic period 0.5~$\mu$m in the [011] direction.

\begin{figure}
\scalebox{0.8}{\includegraphics[width=\columnwidth]{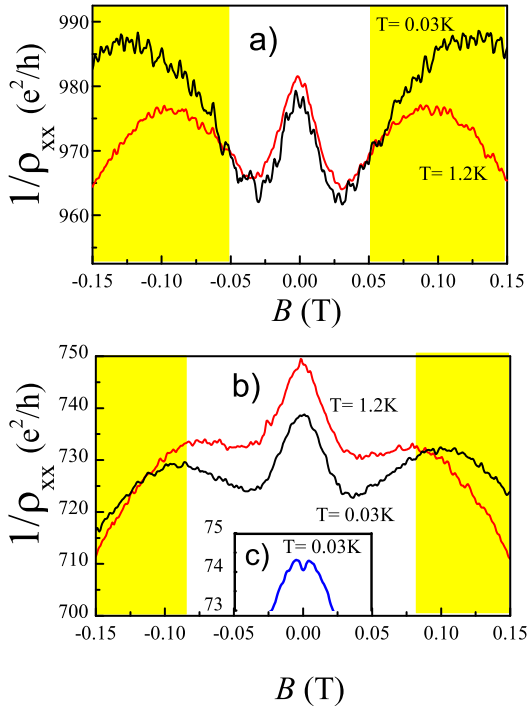}}
\caption{Dependence of $1/\rho_{xx}$ on magnetic field in sample 1 with Hall bar along [011] at different temperatures for the following electron densities: (a) $5\times 10^{11}$~cm$^{-2}$; (b) $4.2\times 10^{11}$~cm$^{-2}$; and (c) $1.9\times 10^{11}$~cm$^{-2}$.}
\label{fig3}
\end{figure}

In Fig.~\ref{fig3}, we show the inverse longitudinal resistance $1/\rho_{xx}$ as a function of magnetic field for the Hall bar along the [011] direction at different electron densities and temperatures. In the range of weak magnetic fields, over the unshaded area, the behavior is similar to that expected for the transition between the weak localization and anti-localization with a distinction that the amplitude of the effects is two orders of magnitude larger than the scale of the quantum corrections and the ``anti-localization'' peak amplitude is temperature independent in the range of temperatures studied. Moreover, as the electron density is decreased, there appears a behavior that can be interpreted as the suppression of the weak localization (Fig.~\ref{fig3}(c)).

\begin{figure}
\scalebox{0.8}{\includegraphics[width=\columnwidth]{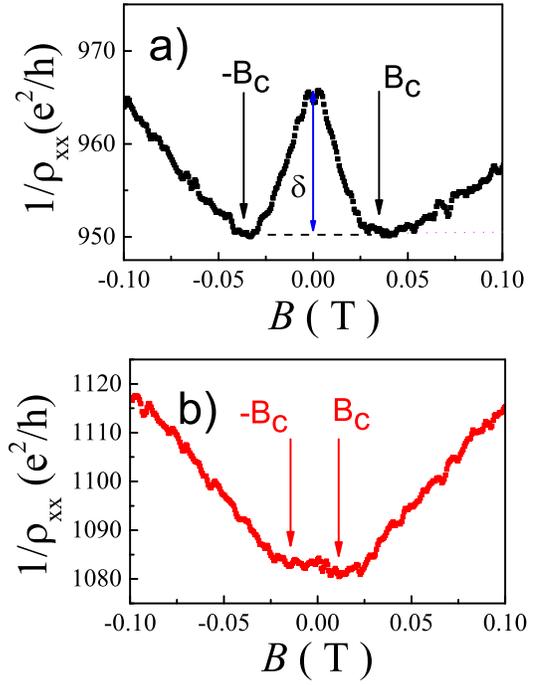}}
\caption{Dependence of $1/\rho_{xx}$ on magnetic field at electron density $5\times 10^{11}$~cm$^{-2}$ and temperature $\approx 0.03$~K for two samples: (a) sample 3 with Hall bar along [011]; and (b) sample 2 with Hall bar along [01-1]. The half-width $B_c$ and amplitude $\delta$ of the peak are indicated.}
\label{fig4}
\end{figure}

In Fig.~\ref{fig4}, we show the inverse magnetoresistance for two Hall bars with different orientations at the same electron density and temperature. For the [01-1] orientation in Fig.~\ref{fig4}(b), the $B=0$ peak is much less pronounced and the adjacent minima, marked by arrows, correspond to smaller values of the magnetic field as compared to the other orientation in Fig.~\ref{fig4}(a). The ratio of the minimum magnetic fields is approximately equal to 3, which corresponds to the ratio of the periods of the ridges in two perpendicular directions, indicating the classical origin of the inverse magnetoresistance maximum in zero magnetic field. Note that the mean free path of electrons is $l\approx 5.5$~$\mu$m at density $5\times 10^{11}$~cm$^{-2}$ in this electron system and the cyclotron radius is $R_c\approx 3.9$~$\mu$m at field 0.03~T.

\begin{figure}
\scalebox{0.6}{\includegraphics[width=\columnwidth]{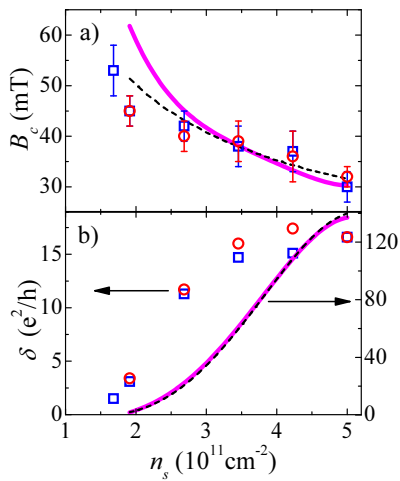}}
\caption{Half-width and amplitude of the inverse magnetoresistance peak as a function of electron density in sample 1 with Hall bar along [011] at temperatures $\approx 0.03$~K (squares) and 1.2~K (circles). The calculation corresponds to Eqs.~(\ref{eq1}-\ref{eq4}) at $V_0=2.4$~mV (dashed lines) and Eq.~(\ref{eq5}) at $V_0=2.2$~mV (solid lines).}
\label{fig5}
\end{figure}

The half-width $B_c$ and amplitude $\delta$ of the inverse magnetoresistance peak versus electron density are shown in Fig.~\ref{fig5} for the Hall bar along the [011] direction. With decreasing electron density the peak broadens and its amplitude decays drastically.

The experimental behavior of the magnetoresistance can be explained in the spirit of Refs.~\cite{ives,weiss} where a periodic potential relief was introduced into the sample and a qualitatively similar magnetoresistance was observed in weak magnetic fields. Let us consider a model potential relief $eV_0\cos(2\pi x/a)$, where the relative modulation $\varepsilon=eV_0/2E_F\ll 1$, $E_F$ is the Fermi energy, and $a$ is the modulation period along the $x$-axis. As was shown in Ref.~\cite{ives}, in such a potential there arise open electron orbits in weak perpendicular magnetic fields. The turning points of the open electron orbit, $x_0$ and $x_1$ ($x_0>x_1$), for electrons moving mainly along the $y$-axis obey the condition

\begin{eqnarray}
&&\frac{x_0-x_1}{R_c}=\frac{eV_0}{2E_F}\left(\cos\frac{2\pi x_1}{a}-\cos\frac{2\pi x_0}{a}\right),\nonumber\\
&&x_0-x_1\leq a.
\label{eq1}
\end{eqnarray}
At a point $x$ between the minimum and maximum turning points $x_{\text{1min}}\leq x\leq x_{\text{0max}}$, there are two regions on the Fermi semicircle: within some angle $\beta(x)$ near the $y$-axis, the open electron orbits are realized in a magnetic field, whereas at higher angles the closed electron orbits are the case. The angle $\beta(x)$ is determined by the orbit with turning points $x_{\text{1min}}$ and $x_{\text{0max}}$

\begin{eqnarray}
\cos\frac{\beta(x)}{2}&=&1+\frac{x-x_{\text{0max}}}{R_c}\nonumber\\
&+&\varepsilon\left(\cos\frac{2\pi x}{a}-\cos\frac{2\pi x_{\text{0max}}}{a}\right).
\label{eqbeta}
\end{eqnarray}
The maximum value of $\beta$ attained in the limit of zero magnetic field is equal to $\beta_{\text{max}}\simeq 4\varepsilon^{1/2}$. With increasing magnetic field the angle $\beta$ decreases until the open orbits disappear at some magnetic field. As one approaches the turning points, $\beta$ decreases and zeroes at $x_{\text{1min}}$ and $x_{\text{0max}}$.

It is easy to demonstrate the role of the open electron orbits to form the magnetoresistance. In an electric field $E_y$, the current from the open orbits in a strip with width $a$ is equal to

\begin{eqnarray}
I_y&=&2\sigma_0\gamma aE_y,\nonumber\\
\gamma&=&\frac{1}{a}\int\limits_{x_{\text{1min}}}^{x_{\text{0max}}}\frac{\beta(x)}{\pi}dx,
\label{eqIy}
\end{eqnarray}
where $\sigma_0$ is the conductivity in the absence of modulation and magnetic field. This leads on average to an increase in the conductivity $\sigma_{yy}$ by $(\Delta\sigma_{yy})_1=2\sigma_0\gamma$. At the same time, the electrons in the open orbits do not contribute to the conductivity due to the closed orbits, leading to a negative correction $(\Delta\sigma_{yy})_2=-2\sigma_0\gamma(1+\omega_c^2\tau^2)^{-1}$, where $\omega_c$ is the cyclotron frequency and $\tau$ is the relaxation time. The entire component $\sigma_{yy}$ of the conductivity tensor is written

\begin{equation}
\sigma_{yy}=\sigma_0(1+\omega_c^2\tau^2)^{-1}+(\Delta\sigma_{yy})_1+(\Delta\sigma_{yy})_2.
\label{eq4}
\end{equation}
Provided $\gamma\ll 1$, the component $\sigma_{xx}$ is not changed by the periodic potential relief and is equal to $\sigma_{xx}=\sigma_0(1+\omega_c^2\tau^2)^{-1}$. The non-diagonal components equal $|\sigma_{yx}|=|\sigma_{xy}|=\sigma_0(1+\omega_c^2\tau^2)^{-1}\omega_c\tau(1-\gamma)$. The above relations allow the analytic calculations of the magnetoresistance modified by open orbits.

We have also performed the numerical calculations of the magnetoresistance for the same model potential relief, using the relation based on the Kubo formula \cite{kubo,goran}:

\begin{equation}
\sigma_{ij}=\frac{e^2m^*}{\pi\hbar^2}<\int\limits_0^\infty dt v_i(0)v_j(t)\exp(-t/\tau)>,
\label{eq5}
\end{equation}
where $m^*$ is the effective electron mass (for our case $m^*=0.035m_e$, here $m_e$ is the free electron mass), $v_i$ and $v_j$ are the electron velocity components, and the angle brackets denote averaging over all orbits passing through the point $x$ and over all points within the strip width $a$.

The calculated magnetic field dependence of $1/\rho_{xx}$ is qualitatively similar to that shown in the unshaded area in Fig.~\ref{fig3}(a, b). The detailed comparison of calculation and experiment is represented in Fig.~\ref{fig5}. The calculated half-width of the inverse magnetoresistance peak as a function of electron density is in satisfactory agreement with the experiment assuming that the amplitude $V_0$ is independent of electron density and the period $a$ is equal to $a=0.5$~$\mu$m (Fig.~\ref{fig5}(a)). The weak dependence of $V_0$ on electron density indicates that the main reason for the potential relief is the residual modulation of In content in the quantum well rather than the ridges on the quantum well surface in Fig.~\ref{fig2}. The amplitude of the inverse magnetoresistance peak versus electron density, calculated for the same values $V_0$ and the experimental relaxation time, is in qualitative agreement with the experiment (Fig.~\ref{fig5}(b)). The quantitative discrepancy is a consequence of the too idealized model, which is similar to Ref.~\cite{gus}. Indeed, the real potential relief is unlikely to be close to a harmonic one.

The behavior of the magnetoresistance at fields above $B_c$ is more complicated. As seen from Fig.~\ref{fig3}(a, b), with increasing $B$ the value $1/\rho_{xx}$ passes a minimum and increases in some region of magnetic fields. The initial interval of the increase may be caused by the disappearance of the open orbits. However, the pronounced temperature dependence of $1/\rho_{xx}$ is observed in this region of magnetic fields, which cannot be explained by classical effects. Moreover, in the shaded area in Fig.~\ref{fig3}(a, b), the derivative $d\rho_{xx}/dB$ changes its sign again, the temperature dependence of the magnetoresistance being still present. Note that the quantum oscillations arise in a magnetic field $\approx 0.7$~T in our samples so that the shaded area corresponds to the intermediate regime in which the quantum effects can already be significant despite the strong overlap of quantum levels.

In summary, we have observed the magnetoresistance of the 2D electron gas in InGaAs/InAlAs quantum wells which is qualitatively similar to that expected for the weak localization and anti-localization but its quantity exceeds significantly the scale of the quantum corrections. The calculations show that the obtained data can be explained by the classical effects in electron motion along the open orbits in a quasiperiodic potential relief manifested by the presence of ridges on the quantum well surface. The results indicate that the likely origin for the potential relief is the residual modulation of In content in the quantum well.

We are grateful to E.~V. Deviatov for helpful discussions. This work was supported by bilateral projects CNR-RFBR. The ISSP group was supported by RFBR grants 15-52-78023 and 16-02-00404, Russian Academy of Sciences, and Russian Ministry of Science.

\end{document}